\documentclass[twocolumn,aps,prl,groupedaddress]{revtex4}
\usepackage{graphicx}
\usepackage{xcolor}
\usepackage{amsmath}
\usepackage{enumitem}

\newcommand{\be}{\begin{equation}}
\newcommand{\ee}{\end{equation}}

\newcommand{\bea}{\begin{eqnarray}}
\newcommand{\eea}{\end{eqnarray}}

\newcommand{\p}{\partial}

\newcommand{\lp}{\left(}
\newcommand{\rp}{\right)}

\renewcommand{\Re}{{\rm \, Re\,}}
\renewcommand{\Im}{{\rm \, Im\,}}
\renewcommand{\vec}[1]{{\bf #1}}

\begin{document}
\title{ Linking Spatial Distributions of Potential and Current in Viscous Electronics
}

\author{Gregory Falkovich$^{1}$ and Leonid Levitov$^2$}

\affiliation{$^1$Weizmann Institute of Science, Rehovot 76100 Israel
\\$^2$Massachusetts Institute of Technology, Cambridge, Massachusetts 02139, USA
}

\date{\today}

\begin{abstract}
Viscous electronics is an emerging field dealing with  systems in which strongly interacting electrons behave as a fluid. 
Electron viscous flows are governed by a nonlocal current-field relation which renders the spatial patterns of current and electric field strikingly distinct.
Notably, driven by the viscous friction force from adjacent layers, current can flow against the electric field,  generating negative resistance, vorticity and vortices.
Moreover, different current flows can result in identical potential distributions. 
This sets a new situation where inferring the electron flow pattern from
the measured potentials presents a nontrivial problem.
Using  the inherent relation between these patterns through the complex analysis, here we propose a method for extracting the current flows from potential distributions measured in the presence of a magnetic field.
\end{abstract}

\maketitle
For electron transport in conductors, one can outline two broadly defined scenarios 
depending on
the relative strength of disorder and interactions\cite{gurzhi63,jaggi91,LifshitzPitaevsky_Kinetics,damle97}.
In the disorder-dominated regime one finds 
 ``individualist" behavior of electrons
moving in straight lines like pinballs bouncing among impurities. 
Fast momentum relaxation gives the familiar Ohm's law with current locally proportional to the electric field. 
In the interaction-dominated regime, when particles exchange their momenta at the rates much faster than the disorder collision rates, 
electrons move in a neatly coordinated way, in many ways resembling the flow of viscous fluids.
When the Umklapp and electron-phonon processes are weak, the collisions conserve momentum, which
thus acquires a new role of a collective variable shared among many particles. 
Electron fluids are predicted to feature a host of fascinating  behaviors\cite{andreev2011,sheehy2007,fritz2008,muller2009,mendoza2011,forcella2014,tomadin2014,narozhny2015,principi2015,cortijo2015,lucas2016}. 
Signatures of viscous flows have been seen in ultra-clean GaAs, graphene and PdCoO${}_2$\cite{dejong_molenkamp,bandurin2015,crossno2016,moll2016}.
The viscous fluid picture has also been invoked to describe collective effects in high-energy physics\cite{kovtun2005,son2007b,karsch2008}. 

Current in an electron fluid is locally proportional to momentum density,
but its relation to the electric field is nonlocal since the viscous force is proportional to the  velocity Laplacian. As a result, the two vector fields, electric field and current, can be quite different. Unraveling the relation between them is one of the main challenges of viscous electronics.
In particular, one needs to come up with ways to reconstruct currents from the potentials, 
measurable by a variety of experimental techniques.
As we will see, while the resulting integral relations
are nontrivial, in two dimensions they can be tackled using
a powerful Cauchy-Riemann-type framework of
complex analysis. This provides a direct link between measured potentials and the current flow patterns.

\begin{figure}[t]
\begin{center}
\includegraphics[width=\columnwidth]{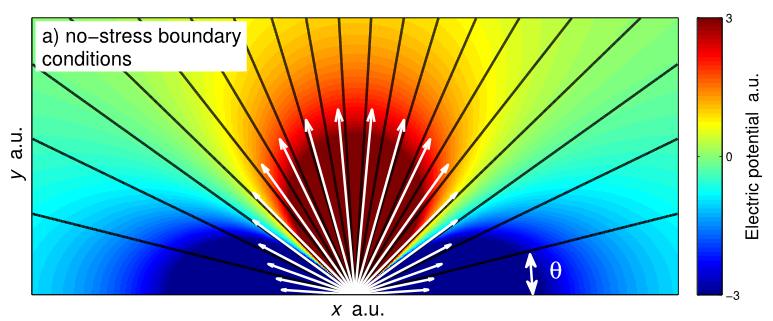} 
\includegraphics[width=\columnwidth]{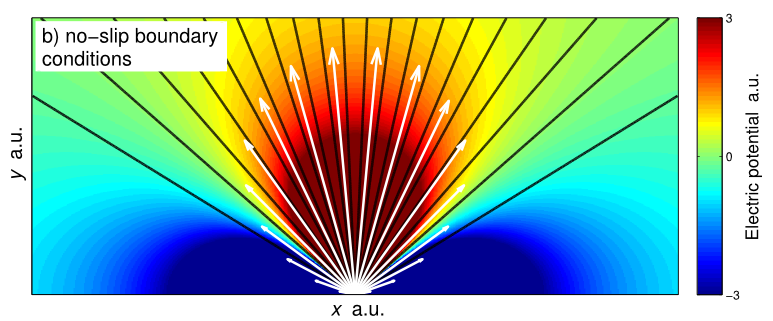} 
\caption{(color online) Streamlines (black) and potential color map for current injected 
through a point in a halfplane,  Eqs.\eqref{eq:psi1psi2},\eqref{no-slip1}.  The velocity is shown by  white
arrows, its magnitude is proportional to the   density of streamlines. 
Boundary conditions: a) no-stress (i.e. shear-stress free); b) no-slip.
}
\label{fig:1}
\end{center}
\end{figure}

We show below that the currents depend not only on the potentials but also, in an essential way, on the boundary conditions.
As a result, identical potential distributions can correspond to {\it totally different flow patterns.}
This surprising behavior
is illustrated in Fig.1 which shows a flow injected into a conducting halfplane through a point-like source at the
edge. For an incompressible flow, charge continuity yields
$\nabla\cdot {\vec j}=ne\nabla\cdot{\vec v}=0$, which can be resolved
by introducing the stream function:
\be\label{eq:stream_function}
\vec v=\vec z\times \nabla\psi=(-\partial_y\psi,\partial_x\psi) .
\ee
The isolines of $\psi$ define
streamlines since their tangent is everywhere parallel to the velocity, see e.g. \cite{F}. Panels (a) and (b) in Fig.1 present the streamlines for the no-stress (i.e. zero shear-stress) and no-slip boundary conditions, respectively. In both cases the streamlines are straight lines pointing outward away from the source. However, the two flows have very different angular distributions,  described by the stream functions
\be\label{eq:psi1psi2}
\psi_{\rm 1}(\theta)=\frac{\tilde I}{4\pi}(\sin 2\theta-4\theta) ,\quad
\psi_{\rm 2}(\theta)=\frac{\tilde I}{2\pi}(\sin 2\theta-2\theta) ,
\ee
where $\theta=\tan^{-1}y/x$ is the polar angle and $\tilde I=I/ne$ is current nondimensionalized with $n$ and $e$, the carrier density and charge. 
The potential map is identical in both cases, taking negative values at the boundary
\cite{LF,Torre2015,bandurin2015}, see Eq.\eqref{no-slip1} and discussion below. 
Interestingly, both flows pictured in Fig.1 have nonzero vorticity $\omega(\vec r)=\nabla\times \vec v$,
however the streamlines do not form loops.
This illustrates that, 
in a departure from a common belief, vortices are not required for negative voltage to occur.

Understanding the current and voltage distributions in   two dimensions is facilitated by the 
complex-variable framework which relates pairs of physical fields by combining them into a single holomorphic function. A  low-Reynolds flow obeys the Stokes equation,  which states that the viscous friction is balanced by the electric force:
\be
\eta\nabla^2\vec v(\vec r)=ne\nabla\phi(\vec r)
.
\label{Visc1}
\ee
Here $\phi(\vec r)$ is the electric potential and $\eta$ is the viscosity.
Combining  \eqref{Visc1} and  \eqref{eq:stream_function}, we see that the vorticity $\omega=\nabla^2\psi=(\partial_x^2+\partial_y^2)\psi$ and $\phi$ form a Cauchy-Riemann pair
\be\label{eq:CR_conditions}
\p_x\omega=(en/\eta)\p_y\phi
,\quad
\p_y\omega=-(en/\eta)\p_x\phi
.
\ee
The quantities $\omega$ and $\phi$ are therefore proportional to the imaginary and real part of a holomorphic function of $z=x+iy$, respectively. 
This behavior is distinct from the Ohmic case $\vec j=en \vec v =-\sigma\nabla\phi$, where
\be
en\vec z\times \nabla\psi=-\sigma\nabla\phi.
\ee
In this case it is the stream function $\psi$ that takes on the role of a Cauchy-Riemann counterpart of  the potential $\phi$.

Extracting the current spatial distributions from those for the potential, 
which are readily measurable by a variety of experimental techniques\cite{yoo97,yacoby99}, can in principle be done by inverting the integral relations \eqref{eq:CR_conditions}. However, instead of facing this formidable computational task, it is more rewarding to seek an alternative way. Here we suggest 
an approach which involves direct measurements rather than indirect computations.
Namely, we propose measuring magnetoresistance in the presence of a classically weak magnetic field, such that the cyclotron radius is much longer than the collision mean free path.
In this case, the Stokes equation  \eqref{Visc1}  acquires an extra term due to the Lorentz force:
$
\eta\nabla^2 \vec v(\vec r)=ne\nabla\phi(\vec r) +ne B\vec v(\vec r)\times\vec z
$
(a hydrodynamic analogy is the Coriolis force in a rotating frame).
Substituting $\vec v(\vec r)=\vec z\times \nabla\psi(\vec r)$ we can write the Stokes equation as
\be
 \eta\nabla^2 \vec v(\vec r)=ne\nabla\phi(\vec r) +ne B\nabla\psi
 .
 \label{mag1}
\ee
Taking the curl of \eqref{mag1} we obtain $(\nabla^2)^2\psi=(\vec v\cdot\nabla)B$. We see that when the magnetic field does not change along the flow, the stream function $\psi$ obeys the bi-harmonic equation identical to that at $B=0$.

Writing \eqref{mag1} as a continuity equation on the momentum flux tensor, we see that  constant $B$ enters only its diagonal (pressure) part:
\be
{\partial\over\partial x_i}\left[ne(\phi+B\psi)\delta_{ik}+\eta\left({\partial v_i\over\partial x_k}+{\partial v_k\over\partial x_i}\right)\right]=0
.
\label{mag2}
 \ee
This allows one to show that constant magnetic field does not affect the boundary conditions on $\psi$ that we consider here. Indeed, the tangential derivative of $\psi$ is completely determined by the incoming/outgoing current. The normal derivative (equal to the tangential velocity) is determined by friction, that is by the continuity across the boundary of the normal flux of tangential momentum 
i.e. the off-diagonal part of the tensor  in Eq.\eqref{mag2}. For example, the no-slip condition (zero tangential momentum) means an infinite friction between fluid and solid boundary, which apparently cannot be affected by the magnetic field. The no-stress condition (zero flux of tangential momentum) takes place when fluid borders the medium which does not support tangential stresses; here again magnetic field does not change the condition.  This is also the case for the mixed boundary condition \eqref{tang} below, where 
the flux of tangential momentum proportional to the tangential velocity can occur at system boundary. 

We therefore conclude that the stream function remains unchanged when a
constant weak $B$ field is applied. 
Repeating the steps that have led us to Eq.\eqref{eq:CR_conditions}, 
we see that in the presence of a $B$ field the Cauchy-Riemann relations are obeyed by the quantities $\omega$
and $\phi+B\psi$.  Under these conditions, the quantity $\phi+B\psi$ must be equal to the potential obtained at $B=0$. 
Therefore, the $\phi$ and $\psi$ dependence on $B$ takes on a very simple form
\be
\phi_{B\ne0}(\vec r)=\phi_0(\vec r)-B \psi_0(\vec r)
,\quad
\psi_{B\ne0}(\vec r)=\psi_0(\vec r)
\ee
where the subscript zero denotes the quantities found at $B=0$.
This relation can be used to obtain the stream function $\psi$ directly from the electric potential measurements. Alternatively, and perhaps more conveniently, $\psi$ can be obtained through antisymmetrization as
\be\label{eq:psi_B}
2 B \psi_0(\vec r)= \phi_{-B}(\vec r)-\phi_{B}(\vec r)
.
\ee
The stream function is a fundamental fluid-mechanic quantity that describes incompressible flows. 
The relation \eqref{eq:psi_B} therefore provides a vehicle that directly relates current flows with the measured potentials.

As one can see from \eqref{eq:CR_conditions}, electric field can only arise in the presence of nonuniform flow vorticity. To understand better the role of vorticity, we recall that viscous friction is determined by the symmetric part of the tensor of velocity derivatives. The vorticity, which is the anti-symmetric part of this tensor, describes rotation of a fluid element as a whole that does not cause friction (e.g. see \cite{F}).
It is vorticity  {\it inhomogeneity} that produces electric field required to balance viscous friction.
The relations \eqref{eq:CR_conditions} imply, in particular, that in irrotational 
viscous flows, wherein $\omega =0$, the electric potential $\phi$ is constant and the electric force vanishes in the bulk. Such ``freely flowing'' currents are described by a velocity potential, $\vec v\propto\nabla\lambda$. 
Potential flows occur when the vorticity
vanishes on the boundaries, in which case it can be shown  to vanish everywhere.
In terms of the electric potential $\phi$ this translates into equipotential i.e. metallic boundaries
 (the fascinating topic of electric field expulsion from viscous charge flows with metallic boundaries  will be discussed elsewhere). In contrast, the potential is not identically constant and the vorticity is nonzero for nonmetallic boundaries, in which case a wide variety of non-trivial current and potential patterns can arise.

A simple and instructive example is provided by viscous flows originating from a point source at the edge of the halfplane $y\geq0$.
Using translational invariance along $x$ we seek the stream function in the form
\be
\psi(x,y)=\int_{-\infty}^\infty \frac{dk}{2\pi} \psi_k(y)e^{ikx}\, .
\label{FT}
\ee
Current $I$ entering through the point $x=0$ gives the boundary condition
$\psi_k(0)=\tilde I/ik$. Another boundary condition must be imposed on the tangential velocity. We consider a general partial-slip boundary condition \cite{Torre2015}
\be
\partial \psi/\partial y =\ell\partial^2\psi/\partial y^2 ,
\label{tang}
\ee
which states that the tangential velocity at the boundary $v_x$ is proportional to the viscous stress $\eta \partial v_x/\partial y$. The limits $\ell\to0$ and  $\ell\to\infty$ give respectively the no-slip and no-stress cases.  Solving the bi-harmonic equation $(\partial_y^2-k^2)^2\psi_k(y)=0$ with these boundary conditions yields
\be
\psi_k(y)={\tilde I\over ik}e^{-|k|y}\left(1+y|k|{1+|k|\ell\over 1+2|k|\ell}\right) .
\label{psik}
\ee
Asymptotic behavior at $x,y\gg\ell$ is governed by $k\ell\ll1 $  and is the same as in the no-slip case while the flow at short distance  at $x,y\ll\ell$ is determined by $k\ell\gg1$  and corresponds to 
the no-stress boundary condition.

It is straightforward to obtain the current and potential in the limiting cases. For the no-stress boundary condition, $\partial^2 \psi/\partial y^2 =0$,  we have
\be
\psi(x,y)={\tilde I\over 2\pi}
\int_{-\infty}^{\infty}e^{ikx-|k|y}{2+y|k|\over 2ik}dk
\ee
Principal-value integration gives $\psi_1(\theta)$ in Eq.\eqref{eq:psi1psi2}.
The vorticity can then be derived as
$\omega=\nabla^2\psi=\tilde I\Im z^{-2}/2$. The potential, obtained from \eqref{eq:CR_conditions},
has a quadrupole form
 \be
\phi(x,y)=\frac{\tilde I\eta}{2ne}
\Re z^{-2}=-\frac{\tilde I\eta}{2ne}{\cos2\theta\over r^{2}}
.
 \label{no-slip1}
 \ee
In the no-slip case,  in a similar vein, we find $\psi_2(\theta)$ in Eq.\eqref{eq:psi1psi2}. 
The quantities $\omega$ and $\phi$, obtained from $\psi_2(\theta)$, have the same form as 
in Eq.\eqref{no-slip1} but are twice larger than in the no-stress case, where there is no edge friction.

Interestingly, while the streamlines are straight lines directed outward from the source in both cases, the actual velocity patterns are quite different (see Fig.1). For a finite $\ell$ the streamlines are more spread-out near the source, curving up at $x,y\simeq\ell$. Both the viscous force and the 
electric force, balancing each other, are now nonzero. 
The electric force has two components:
\be
{\partial\phi\over \partial r}\propto \frac{\cos2\theta}{ r^3},\quad
r^{-1}{\partial\phi\over \partial \theta}\propto \frac{\sin2\theta }{r^3}.
\ee
The radial electric field changes sign at $\theta=\pi/4$: at $\theta>\pi/4$ it pushes the fluid up, whereas at $\theta<\pi/4$ the electric field is
directed {\it towards} the source, balancing the viscous drag from the faster-moving  adjacent layers of the fluid. It is this field that produces the negative voltage at the edge. 

We note that the no-stress case with a straight boundary is degenerate because the boundary condition $\partial^2\psi/\partial y^2=0$
 does not select a unique solution. Indeed,
a general solution $\psi=-2\tilde I\theta+b\sin2\theta$  fits the bill 
for any $b$. In particular, this is the case for the whole-space solution $\psi= -2\tilde I\theta$,  $\phi=0$, $\omega=0$. The degeneracy can be removed by altering the boundary shape or the boundary conditions.
 This is illustrated by Eq.\eqref{psik} which gives a  well-defined solution in the limit $\ell\to\infty$.

Having established that vorticity is necessary for the appearance of electric field inside a viscous charge flow, we now proceed to discuss vortices. 
Let us state upfront that it is important to distinguish the generic features
due to local vorticity from a more specific global pattern of a vortex. Indeed, nonzero vorticity at a point means that an infinitesimal fluid element rotates as it moves. Such motion, however, may 
take place even along perfectly straight streamlines such as those in the flows pictured in Fig.\ref{fig:1}, where vorticity is non-zero since different streamlines have different velocities. Vortices, on the other hand, are defined by closed-loop streamlines, that is they are global rather than local structures.
Accordingly,
unlike the halfplane geometry in Fig.\ref{fig:1}, vortices can be readily produced  in a confined geometry. 
Vortices can be characterized by {\it separatrix} lines which separate the closed and open streamlines.
Below we illustrate this general behavior for a strip of a finite width.

\begin{figure}[t]
\begin{center}
\includegraphics[width=\columnwidth]{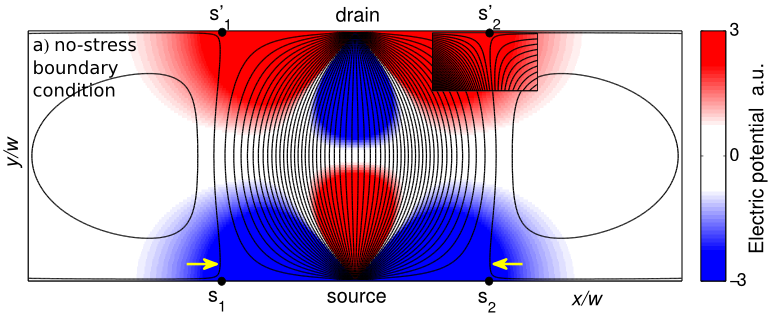} 
\includegraphics[width=\columnwidth]{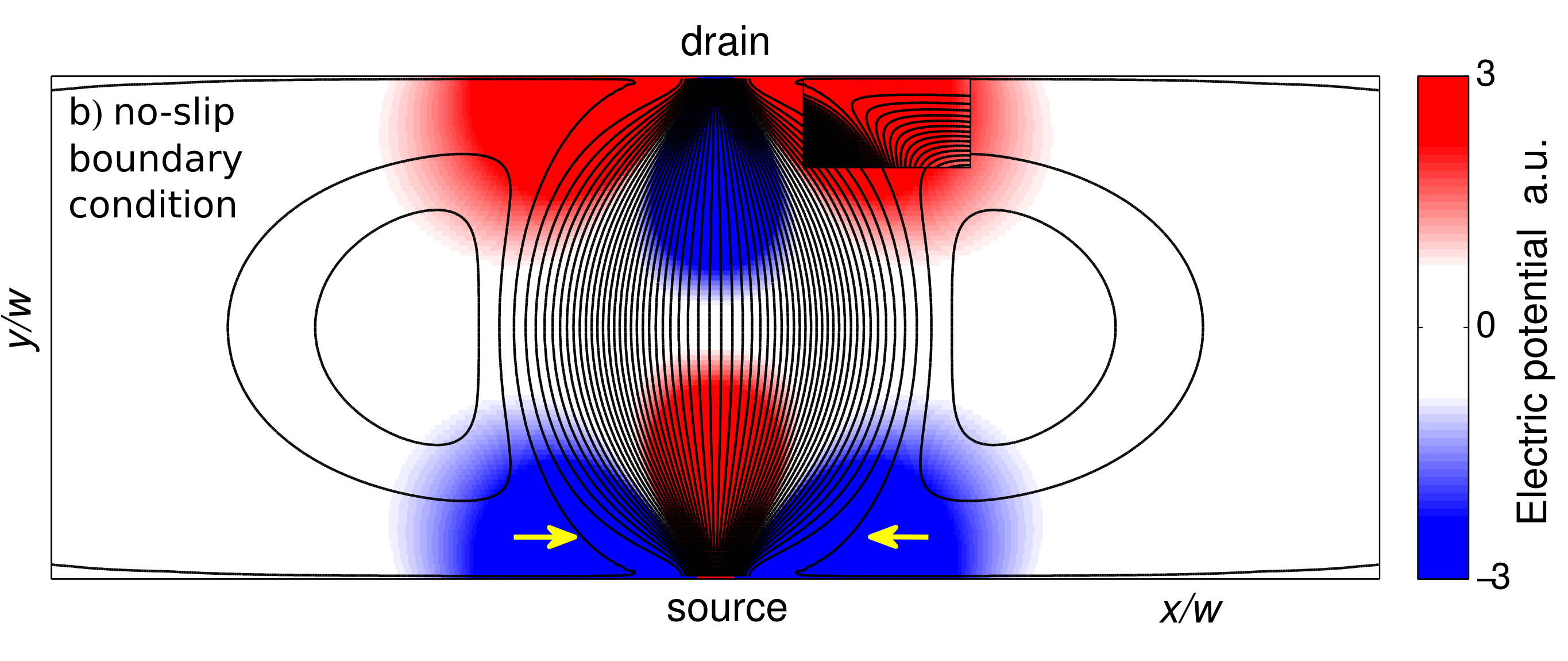} 
\caption{(color online) Current streamlines (black) and potential color map for a flow across the strip.   Arrows mark the streamlines nearest to separatrices. 
Stagnation points are labeled $s_{1,2}$, $s_{1,2}'$. To elucidate the behavior near contacts, two regions are shown with ten times higher density of streamlines. 
Boundary conditions: a) no-stress, b) no-slip.
}
\label{fig:2}
\end{center}
\end{figure}

We start with the no-stress boundary condition and consider the point-like source and drain positioned at $(0,0)$ and $(0,w)$ in the strip $-\infty<x<\infty$, $0<y<w$. 
A solution of the bi-harmonic equation of the form \eqref{FT} with $\psi_k=\tilde I/ik$ and $\partial^2\psi/\partial y^2=0$ at $y=0,w$ reads
\be
\psi(x,y)\!=\!{\tilde I \over 4\pi }\!\!\int\limits_{-\infty}^\infty\!\! {e^{ikx}dk\over i k \cosh\frac{kw}2}
\lp  a  \cosh k\tilde y - k\tilde y \sinh k\tilde y\rp\!,
\label{Sol3}
\ee
where we defined $\tilde y=y-w/2$ and $a =2+\frac{kw}2\tanh\frac{kw}2$.  The streamlines, given by the contours of $\psi$, are pictured in Fig.\ref{fig:2}a. The flow, directed from source to drain along the nominal current path, mimics that in Fig.\ref{fig:1} near each contact. Closed streamlines form a pair of  vortices.

To analyze the separatrices of the flow, we consider the velocity at the boundary $y=0$. Simple algebra yields
\be \label{SP}
v_x(x, 0)=-{\partial\psi\over\partial y}={I \left(2- {\pi x\over w}\coth{\pi x\over w}\right)\over 4 w ne\sinh(\pi x/w)}
.
\ee
At $|x|\ll w$, the velocity is directed away from the source as in a halfplane, $v_x\propto 1/x$. However, $v_x$ is directed {\it towards} the source at $|x|\gg w$,  representing backflow due to vortices. 
We therefore conclude that there are stagnation points at the edge, where $v_x=v_y=0$. At such points, marked $s_1$, $s_2$, $s'_1$, $s'_2$ in Fig.\ref{fig:2}a, two streamlines meet: one directed along the strip edge and another perpendicular to it. The latter represents a separatrix between the source-to-drain streamlines and the vortex streamlines.
The stagnation points are defined by the equation $ \pi x / w =2\tanh(\pi x/ w)$, giving  $x/w=\pm 0.61...$. This is in accord with the flow shown in Fig.~\ref{fig:2}a, where arrows mark the streamlines nearest to the separatrices.

The  potential is obtained by solving Eq.\eqref{Visc1} which gives
\be
\phi(x,y)=\alpha\!\! \int\limits_{-\infty}^\infty \!\! dk e^{ikx} \frac{k \sinh k\tilde y}{\cosh\frac{kw}2}
=\frac{\alpha\pi^2}{w^2}\Re\frac{\cosh \pi z}{\sinh^2\pi z}
,
\ee
where $\alpha=I\eta/\pi (ne)^2$ and   $z=(x+iy)/w$. Amusingly,  this result can also  be obtained
 from the solution for the source and drain in the halfplane, $\phi(z')=\Re\bigl[(z'-1)^{-2}-(z'+1)^{-2}\bigr]$, by mapping it onto the strip.
Both the potential and the flow, taken near
each contact, mimic those found for a point source in the halfplane.

The topology of the flow can change drastically upon altering the boundary conditions. 
 As we now show, the flow found for the no-stress case undergoes a global change
upon switching to the no-slip boundary conditions. This behavior is a manifestation of the fundamental nonlocality of  viscous flows discussed above.
The stream function for the no-slip case is of the form \cite{LF}:
\be\label{Sol1}
\psi(x,y)=\! {\tilde I\over 2\pi}\int\limits_{-\infty}^\infty\! {dk\over i k}e^{ikx} \frac{c_1\cosh k\tilde y-c_2 k\tilde y\sinh k\tilde y}{kw + \sinh kw}
\ee
where $c_1=kw\cosh\frac{kw}2+2\sinh\frac{kw}2$, $c_2=2\sinh\frac{kw}2$. From Fig.\ref{fig:2}b it may appear that the streamlines form radial patterns near contacts identical to those in Fig.\ref{fig:2}a, with
$-\tilde I/2<\psi<\tilde I/2$. However, a closer inspection reveals
additional streamlines  corresponding to the boundary values $\psi=\pm\tilde I/2$. 
These streamlines leave the contacts
horizontally and then curve inward. 
Their form can be obtained explicitly by evaluating $\psi$ in the domain $y\ll x\ll w$. Treating $kw$ as a large parameter, we write  
\bea\nonumber
\pi \psi(x,y)/\tilde I & \approx&  \arctan(x/y)+xy/(x^2+y^2)+2xy/w^2
\\ &&
\approx \pi/2-2y^3/3x^3+3xy^2/w^3. 
\eea
The terms linear in $y/x$ cancel, which allows for a second streamline with the same $\psi$ value as at the edge, $\psi(x,y)=\tilde I/2$. This line, described by $y=9x^4/2w^3$ at small $y$, is a separatrix between 
the source-to-drain streamlines and the vortex streamlines. 
This is illustrated in  Fig.\ref{fig:2}b where arrows mark the streamlines nearest to the separatrices.
The vortex streamlines fill the space between the separatrix and the strip edge,
extending arbitrarily close to the contacts. 

To confirm in a different way that the streamlines below the separatrix turn around without reaching
the source,  we analyze the
velocity component $v_x$.   Evaluating $v_x=-\frac{\p\psi}{\p y}$ and, for $ y\ll x\ll w$, approximating
\be
\!\!\int\limits_{ 0}^\infty  {k\over 2} dk \sin kx (ye^{-ky}-w\sinh ky e^{-kw})
\approx {y^2\over x^3}-{xy\over w^3}
,
\label{Sol2}
\ee
we see that the horizontal velocity reverses its sign at the `demarcation' line $y=x^4/w^3$  (lying below the separatrix) which means that upon crossing this line the streamlines turn around. Below this line, the second term in  \eqref{Sol2} dominates, making the flow along the edge directed towards the contact.
In the limit $w\to\infty$, when the strip turns into halfplane, 
the demarcation line disappears. 
In this case there are no closed streamlines and no backflow.

To understand the role of sample shape, we replace the infinite strip by a rectangle of size $w\times L$. 
In this case, one must replace $\int dk \sin kx \rightarrow\sum_n\sin k_nx$ with $k_n=\pi n /2L$. For $L\gg w$ the flow remains unchanged. When $L\lesssim w$, the last integral in \eqref{Sol2} is determined by large $k$ and is insensitive to this change so long as $x,y\ll L$. The first integral in \eqref{Sol2} is replaced by the first term of the sum, $-k_1^3 xye^{-k_1w}$. Henceforth, the separatrix survives for any $L$ but moves closer to the edge as $L/w$ decreases: $y\propto x^4L^{-3}\exp(-\pi w/2L)$. 
Therefore, somewhat counterintuitively, the vortices persist at any aspect ratio $w/L$.


Potential distribution, obtained directly from Eq.\eqref{Visc1}, looks qualitatively similar to that found for the no-stress case, see Fig.\ref{fig:2}b.
It changes sign twice on the nodal lines that come out of the contacts at the $\pm\pi/4$ angles, like in a halfplane, that is the behavior of $\phi(x,y)$ does not reflect the presence of the separatrix and backflow.
True, when one runs the source-drain current across the strip, 
there is always a backflow along the edges. 
Yet, this backflow (however spectacular by itself) is of little relevance for the negative voltage that exists even in a halfplane with no vortices. Likewise, the voltage singularities near the contacts are due to converging/diverging streamlines and have nothing to do with vortices or separatrix.

The above discussion illustrates  general implications of the nonlocal current-field relations. Most important, we have shown that it is the negative electric field rather than the backflow that 
 is a true universal signature 
 of the viscous electronics. While the negative field is inherently 
related to the vorticity of current flow, it requires neither backflow nor vortices. 
Further, there is no one-to-one relation between the spatial distributions of currents and potentials,
making it non-trivial to infer the current flow from the measured potential.
An answer is provided by application of a weak magnetic field, which effects a change in the potential distribution proportional to the current  stream function. This 
opens door to direct measurements of viscous electron flow patterns 
by the well-developed charge and potential sensing techniques\cite{yoo97,yacoby99}.

\end{document}